%Paper: astro-ph/9401032
%From: tsvi@shemesh.fiz.huji.ac.il (Tsvi Piran)
%Date: Mon, 17 Jan 94 15:44:56 +0200
%Date (revised): Mon, 17 Jan 94 22:27:35 +0200

%\hoffset=0.7cm \voffset=0.2cm
%\hoffset 0.5truein
\hsize 6.5truein
\vbadness=10000
%
% The following statements redefine the basic fonts
% to be magnified by a factor 1.2
%

\font\sss=cmssq8 scaled 1000
\font\bf=cmbx10 scaled 1200
\font\bb=cmbx10 scaled 1920

\font\rs=cmr8 scaled 1200
\font\it=cmti10 scaled 1200

\font\sl=cmsl10 scaled 1200
\font\sc=cmcsc10 scaled 1200
\font\tenrm=cmr10 scaled 1200
\font\sevenrm=cmr7 scaled 1200
\font\fiverm=cmr5 scaled 1200
\font\teni=cmmi10 scaled 1200
\font\seveni=cmmi7 scaled 1200
\font\fivei=cmmi5 scaled 1200
\font\tensy=cmsy10 scaled 1200
\font\sevensy=cmsy7 scaled 1200
\font\fivesy=cmsy5 scaled 1200

\font\tenbf=cmbx10 scaled 1200
\font\sevenbf=cmbx7 scaled 1200
\font\fivebf=cmbx5 scaled 1200
\font\tensl=cmsl10 scaled 1200
\font\tentt=cmtt10 scaled 1200
\font\tenit=cmti10 scaled 1200
\catcode`\@=11
\textfont0=\tenrm \scriptfont0=\sevenrm \scriptscriptfont0=\fiverm
\def\rm{\fam\z@\tenrm}
\textfont1=\teni \scriptfont1=\seveni \scriptscriptfont1=\fivei
\def\mit{\fam\@ne} \def\oldstyle{\fam\@ne\teni}
\textfont2=\tensy \scriptfont2=\sevensy \scriptscriptfont2=\fivesy
\def\cal{\fam\tw@}
\textfont3=\tenex \scriptfont3=\tenex \scriptscriptfont3=\tenex
\newfam\itfam \def\it{\fam\itfam\tenit} % \it is family 4
\textfont\itfam=\tenit
\newfam\slfam \def\sl{\fam\slfam\tensl} % \sl is family 5
\textfont\slfam=\tensl
\newfam\bffam \def\bf{\fam\bffam\tenbf} % \bf is family 6
\textfont\bffam=\tenbf \scriptfont\bffam=\sevenbf
\scriptscriptfont\bffam=\fivebf
\newfam\ttfam  % \tt is family 7
\textfont\ttfam=\tentt
\catcode`\@=12
%
% This ends font redefinitions
%
\rm

% Definitions added in my version

\abovedisplayskip=30pt plus 4pt minus 10pt
\abovedisplayshortskip=20pt plus 4pt
\belowdisplayskip=30pt plus 4pt minus 10pt
\belowdisplayshortskip=28pt plus 4pt minus 4pt
\def\folio{\ifnum\pageno=1\nopagenumbers\else\number\pageno\fi}

\def\--{\! - \!}

\hfuzz=10pt \overfullrule=0pt
\vsize 9truein
%\hsize 6truein
\baselineskip=14truept

\parindent 30pt \parskip 6pt

\def\blankline{\par\vskip \baselineskip}
\def\twocol#1{\halign{##\quad\hfil &##\hfil\cr #1}}

 \mathcode`*="002A

\def\_{\vrule height 0.8pt depth 0pt width 1em}

\newbox\grsign \setbox\grsign=\hbox{$>$} \newdimen\grdimen \grdimen=\ht\grsign
\newbox\simlessbox \newbox\simgreatbox
\setbox\simgreatbox=\hbox{\raise.5ex\hbox{$>$}\llap
     {\lower.5ex\hbox{$\sim$}}}\ht1=\grdimen\dp1=0pt
\setbox\simlessbox=\hbox{\raise.5ex\hbox{$<$}\llap
     {\lower.5ex\hbox{$\sim$}}}\ht2=\grdimen\dp2=0pt
\def\simgreat{\mathrel{\copy\simgreatbox}}
\def\simless{\mathrel{\copy\simlessbox}}

\def\dot#1{\vbox{\baselineskip=-1pt\lineskip=1pt
     \halign{\hfil ##\hfil\cr.\cr $#1$\cr}}}
\def\ddot#1{\vbox{\baselineskip=-1pt\lineskip=1pt
     \halign{\hfil##\hfil\cr..\cr $#1$\cr}}}

\def\cfalh{\par\vfil\eject \vskip -12pt \moveleft 0.5in\vbox{
     \twocol{{\bb Center for Astrophysics}\hbox to 1.5in{} &\cr
     {\sss 60 Garden Street} & {\sss Harvard College Observatory} \cr
     {\sss Cambridge, Massachusetts 02138} & {\sss Smithsonian
          Astrophysical Observatory}\cr}}\par\blankline}

\def\listitem{\par \hangindent=50pt\hangafter=1
     $\ $\hbox to 20pt{\hfil $\bullet$ \hfil}}
\def\date#1{\par\hbox to \hsize{\hfil #1\qquad}\par}

\def\ref#1{$^{#1}$}
\def\title#1\endtitle{\par\vfil\eject
     \par\vbox to 1.5in {}{\bf #1}\par\vskip 1.5in\nobreak}
\def\author#1\endauthor{\par{\sc #1}\par\blankline}
\def\institution#1\endinstitution{{\rs #1}}

\def\section#1\endsection{\par\vfil\eject{\bf #1}\par\vskip 12pt\nobreak\rm}
\def\subsection#1\endsubsection{\vskip 14pt plus 50pt {\rm #1}\par
     \nobreak\blankline\nobreak\rm}

%\baselineskip=28truept
%\doublespace
%\hbox{ }
%\vskip 0.5in
\centerline{\bf MERGING NEUTRON STARS I:}
\vskip .1in
\centerline{\bf INITIAL RESULTS FOR COALESCENCE OF NON-COROTATING SYSTEMS}
\vskip .3 in
\centerline {\bf M. B. Davies}
\centerline{Theoretical Astrophysics, 130-33}
\centerline{California Institute of Technology}
\centerline{Pasadena, CA 91125}
\vskip .2 in
\centerline{\bf W. Benz}
\centerline{Steward Observatory}
\centerline{University of Arizona}
\centerline{Tucson, AZ 85721}
\vskip .2in
\centerline{and}
\vskip .2in
\centerline{\bf T. Piran$^1$, F. K. Thielemann}
\centerline{Harvard-Smithsonian Center for Astrophysics}
\centerline{60 Garden Street}
\centerline{Cambridge, MA 02138}
\vfill
\centerline{Submitted to}
\centerline{\it The Astrophysical Journal}
\centerline{September, 1993}
\vskip .1in
\vfill
 $^1$Permanent address: Racah Institute for Physics, The
Hebrew University, Jerusalem, Israel 91904.

\eject
%\baselineskip=28truept
\def\etal{{\it et. al.}}
\centerline {\rm ABSTRACT}
\nobreak
\medskip
We present 3D Newtonian simulations of the coalescence of
two neutron stars, using a Smoothed Particle Hydrodynamics (SPH) code.
We begin the simulations with the two stars in a hard, circular binary,
and have them spiral together as angular momentum is lost through gravitational
radiation at the rate predicted by modeling the system as two point masses.
We model the neutron stars as hard polytropes ($\gamma=2.4$) of equal
mass, and investigate the effect of the initial spin of the two
stars on the coalescence. The process of coalescence, from initial contact to
the formation of an axially symmetric object, takes only a few orbital
periods. Some of the material from the two neutron stars is shed, forming
a thick disk around the central, coalesced object. The mass of this disk is
dependent on the initial neutron star spins; higher spin rates resulting
in greater mass loss, and thus more massive disks. For spin rates
that are most likely to be applicable to real systems, the central
coalesced object has a mass of $2.4M_\odot$, which is tantalizingly
close to the maximum mass allowed by any neutron star equation of state
for an object that is supported in part by rotation.
Using a realistic nuclear equation of state, we estimate the temperature
of the material after the coalescence. We find that the central object
is at a temperature of $\sim 10$MeV, whilst the disk is heated by shocks
to a temperature of 2-4MeV.

\bigskip
\centerline {1. INTRODUCTION}
\nobreak
\medskip

It has been suggested that $\gamma$-ray bursts arise at cosmological
distances in the merger of binaries consisting of either two neutron
stars, or a black hole and a neutron star (Eichler \etal$\,$ 1989,
Piran, 1990, Paczy\'nski, 1991, Piran \etal$\,$ 1992, Narayan,
Paczy\'nski \& Piran 1992).  Two different processes may provide the
electromagnetic energy for the bursts: neutrino-antineutrino
annihilation into electron-positron pairs during the merger, and
magnetic flares generated by Parker instabilities in a postmerger
differentially-rotating disk.  Merging binary neutron stars are also a
source of gravitational radiation, whose waveform and amplitude can be
predicted with reasonable confidence (see e.g. Lincoln \& Will, 1990;
Cutler, \etal$\,$, 1993) and are prime candidates for detection by
LIGO (Schutz, 1986, Thorne, 1987; Abramovici \etal$\,$ 1992).  It has
been estimated that $\sim 3$ such mergers will occur in a year within
200 Mpc (Narayan, Piran \& Shemi 1991, Phinney 1991). These events
should be observable by the Laser Interferometer Gravitational Wave
Observatory (Abramovici \etal$\,$ 1992).  A merger of two neutron
stars may lead to the ejection of neutron-rich material that may be a
good site for the production of r-process elements.  It is possible
that such mergers account for all heavy r-process material in the
Galaxy (Eichler \etal$\,$ 1989).

The production of binaries containing two neutron stars has been
discussed in the literature (see for examples, Narayan, Piran \& Shemi
1991; van den Heuvel 1993). For completeness, the scenario is outlined
here.  Begin with two massive stars in a binary. The primary -- the
more massive of the two stars -- will evolve first, forming a neutron
star in a supernova. If the binary is tight enough, and the secondary
sufficiently massive, the binary may not be disrupted by this violent
event. Mass transfer between the secondary and neutron star will ensue
as the former evolves beyond the main sequence, the star either
filling its Roche lobe, or losing mass via a wind. At this stage in
its evolution, the system will manifest itself as a massive X-ray
binary.  If the secondary were to evolve to a supernova without
significant mass loss and if the explosion is symmetric, the binary
would be assured of disruption, as the mass lost from the exploding
star would be much greater than that remaining in the two neutron
stars. However, we may speculate upon the following panacea:
because the bloated secondary is more massive than the neutron star,
mass transfer from the former will be unstable, if the mass transfer
rate is large enough, bringing the two stars closer together until
they form a common envelope system. Such a system represents an
efficient way to remove the nefarious excess matter in the secondary's
envelope.  By finely tuning the separation of the neutron star and
core of the secondary at the beginning of the common envelope phase,
we are able to speculate that one may remove a large fraction of the
gaseous envelope, leaving the core of the secondary and the neutron
star in a tight binary.  When the core finally explodes, the relative
mass loss from the system may thus have been reduced sufficiently to
avoid disruption.  Alternatively, the supernova explosion might be
assymetric leaving the neutron stars in an eccentric orbit. It is
unclear how likely the binary is to remain intact. Estimates based on
the current known neutron stars binaries suggest that this happens in
about one percent of the cases (Narayan, Piran \& Shemi, 1991).  The
fragility of such systems is evidenced by the eccentricities of the
four known neutron star binaries, suggesting these systems were close
to disruption when they were formed. It should be noted that
systems formed in this manner will have circularized before the
two neutron stars come into contact (C. Cutler, private communication).

Two neutron stars in a binary will spiral together as angular momentum
is lost via the emission of gravitational radiation.  The timescale
for merger of two neutron stars, with $M_1=M_2=1.4M_\odot$, is less
than $10^{10}$ years if the initial separation satisfies $d
\simless 5 R_\odot$ (Iben and Tutukov 1984).  Recent work (Kochanek
1992; Bildsten \& Cutler, 1992; Lai, Rasio \& Shapiro, 1993)
has shown that the viscosity of the
neutron-star fluid is unlikely to be sufficient to spin-up the stars
to achieve tidal-locking as the two stars spiral together.  One
therefore has to ask how the merger process will change as a function
of neutron star spin.

Hydrodynamic simulations of neutron-star mergers are a formidable
task.  Beyond the 3D hydrodynamic calculations it should include
general relativitistic effects, employ a realistic equation
of state, contain neutrino transport and neutrino cooling, and possible
nuclear reactions.
So far there have been several attempts to address various aspects of
the problem. Most of these attempts focused on the gravitational
radiation emmission (Oohara \& Nakamura, 1989, 1990; Nakamura \&
Oohara 1989, 1991; Shibata, Nakamura \& Oohara, 1992).  These were
Newtonian calculations using a finite difference code,
a polytropic equation of state and
including a gravitational radiation backreaction formula.
Rasio \& Shapiro (1992)
focused on the hydrodynamic evolution of the rotating
core and began with the two neutron stars
corotating at contact.
Here we present a study that focuses on the
thermodynamics and nuclear physics of the coalescence.  As was
mentioned above, the low viscosity of the neutron-star material leads
us to consider the merger of {\it non-corotating} neutron stars.  In
order to achieve a realistic estimate of the radial velocity at the
moment that the neutron star come in to contact we begin with the two
neutron stars being far apart and follow the spiral-in phase. We
follow the evolution until the neutron stars merge and we examine both
the matter distribution and the thermal properties of the system after
the two neutron stars merge. Our goal is to estimate the thermal
conditions in the coalesced core and the disk that forms around it
after the two neutron stars merge, and to study the possible
subsequent nuclear processes and neutrino radiation processes.  As a
first stage we do not attempt to include relativistic effects, or
even the complete equation of state in these calculations. At this
stage, we perform Newtonian calculations with a polytropic equation of
state, with a gravitational radiation backreaction formula. We
employ, however, a realistic equation of state to estimate the
thermodynamic conditions during the computations. We compare the
conditions prevailing in our simulations to those required to produce
a gamma-ray burst by neutrino-antineutrino annihilation (Eichler
\etal$\,$ 1989) or by accretion of the disk onto the compact core
(Narayan, Paczy\'nski \& Piran 1992).

The numerical methods applied to this problem and the initial
conditions used are described in section 2. The results of our
simulations are presented in section 3 and in section 4 we discuss the
various physical processes that could take place in this system.  The
implications of these results to the question of formation of
gamma-ray bursts and to r-process nucleosynthesis are summarized in
section 5.

\bigskip
\centerline{2. NUMERICAL METHODS AND INITIAL CONDITIONS}
\nobreak
\medskip

We performed a series of simulations of the merger of two neutron
stars of equal mass, using a 3D Smoothed Particle Hydrodynamics (SPH)
code, based on one used previously to study collisions between stars
(see for example, Benz \& Hills 1992). Being a Langrangian particle
code, SPH is well suited to this problem.  We have no need for a
computational box (gravitational forces being computed using a tree),
hence we do not waste computational resources simulating the evolution
of the voids between the neutron stars as they spiral together.

Since we are mostly interested in the thermodynamic conditions of the
material we are using a Newtonian code, rather than a general
relativistic one. The only general relativistic effect which we
include is the addition of the gravitational radiation backreaction
force, which we discuss below.
We modelled the cold neutron stars as hard polytropes.
We employed a realistic nuclear equation of state developed by Lattimer
\& Swesty (1991), to compute pressure as a function of density for cold
material
at nuclear densities with a compressibility $K=180$ MeV,
and obtained $\gamma=2.4$
for a polytropic approximation of the equation of state. We checked aposteriori
that the Lattimer
\& Swesty equation of state is well approximated by a polytrope for
the density range that appears in our calculations (except for a small
amount of mass in the very low density region). In view of the large
uncertainty in the equation of state of nuclear matter we believe that
this is a reasonable first step.  We later make use of the Lattimer \& Swesty
equation of state to estimate the temperature from the internal energy and
density at the final stages of the computation.  To construct the
initial neutron-star model, we solve the Lane-Emden equation for
$\gamma=2.4$, and we construct a polytrope with the required density
profile using the method described in Davies, Benz \& Hills (1992).

We began each simulation with the two stars $5 R_{\rm ns}$ apart and
had them spiral in at the rate predicted by treating the system as two
point masses and considering the energy lost through the emission of
gravitational radiation. The change in energy, $E$, and angular
momentum, $J$, is given by (e.g. Shapiro \& Teukolsky 1983)
$$\eqalignno{ {dE\over dt} = \alpha E, & \ \ \ \  { d J\over dt} = - {\alpha
\over 2} J &(1)\cr
{\rm where} \  \alpha = { 64 \over 5} & {G^3 \over c^5} {M^2 \mu
\over a^4}\cr}$$
where $a$ is the binary separation, $M$ the total mass of the system,
and $\mu = M_1 M_2 /(M_1 + M_2)$ is the reduced mass, where $M_1=
M_2=(1/2) M$ is the mass of the two neutron stars.
Using the above equations, we derive the following formulae for the
acceleration due to the emission of gravitational radiation.
$$\eqalign{\ddot x &= - {GM_1 \over 2 r^3} x + {\alpha \over 2 M_1 (\vec v
\cdot \vec r)} \left( E x - {J  \dot y \over 2} \right)\cr
\ddot y &= - {GM_1 \over 2 r^3} y + {\alpha \over 2 M_1 (\vec v
\cdot \vec r)} \left( E y + {J \dot x \over 2} \right)\cr}
\eqno(2)$$
As we apply the same acceleration to each SPH particle within each neutron
star, the circulation of the fluid will be conserved, as the accelerating
field has a zero curl.
This in-spiral force
was turned off once the two stars came into contact. At this stage it
is clear that this approximation breaks down. While this force is a
far cry from a sofisticated gravitational radiation emission
backreaction force (Blanchet, Damour \& Schafer 1990) it provides us
with a simple formula that can be easily added to our Newtonian
calculations
which will provide a realistic radial velocity component at the moment when the
two stars collide.

As the two stars spiral together, a lag angle will
develop between the stars' long-axes and the line joining their
centers-of-mass.
The size of this angle is partially a function of the viscosity
of the neutron star fluid.  As the fluid inside the neutron stars is
extremely inviscid, we expect the figures of the two neutron stars to
remain essentially alligned, until the very last
stages of the in-spiral. However, one might be concerned that the much larger
effective viscosity present in our fluid would lead to
an artificially-large lag angle, with the two stars then receiving torques.
Indeed, in early, lower-resolution runs, the stars were spun
up close to the orbital spin rates.
To circumvent this problem, we ran a second set of
mergers, alligning the two neutron stars every time-step, so that their
long-axes remained parallel. The shift in total
angular momentum due to this procedure was measured and found to be tolerably
small ($\sim 1$ in $10^5$).
In fact, the two higher resolution runs presented here (runs C
and D in Table I) that used this reallignment mechanism show very similar
results to the equivalent runs performed without the reallignment (runs
A and B). In other words, the problem which seemed to require the use
of reallignment largely disappeared when the resolution was increased
in the final runs we performed. Both sets of runs are given here for
completeness.

Each simulation was run for 3.5ms ($\sim 50$ dynamical times)
after the stars came into contact. This
was sufficient time for all mergers to produce a central coalesced
object surrounded by a disk. As we show later, our simplifying
assumptions do not break down at this stage. However, the
evolution of this system on longer time scales is determined by
processes, which we did not include, but address in detail in section 4.

\bigskip
\centerline{3. SPH POLYTROPIC RESULTS}
\nobreak
\medskip

We performed four high-resolution mergers (runs A -- D) and two lower
resolution simulations (runs E and F).
In runs A -- D, we used 4271 particles per neutron star,
in runs E and F, we used 1256 particles per neutron star.
In these six runs we keep all
initial parameters the same except the initial spins of the two
neutron stars.  The spin angular velocities of the neutron stars when
they came into contact is given in Table 1. For run A the
lag angle $\sim 4^\circ$ by the time the two stars come into contact.
For run B, the value is somewhat larger at $\sim 10^\circ$.  Taking
$M_{\rm ns} = 1.4M_\odot$ and $R_{\rm ns} = 10 $km, $\Omega_{\rm
spin} = 1$ (the synchronous spin with the two neutron stars
$2R_{\rm ns}$ apart) corresponds to a spin period of 0.46ms.  Hence in runs A
and C, the two neutron stars have spin periods $\sim 2.5$ms. We
consider this to be a reasonable upper limit to the neutron star spins
for any merging system, as the stars are unlikely to be spun up to
their orbital periods just before contact, owing to the low viscosity
of the neutron star fluid. In runs B and D, we consider the case where
the two neutron stars have negligible spin. We expect essentially all
systems to lie in between these two cases. In runs E and F, we further
investigate the effect of the neutron star spins, by considering cases
where the stars are spinning {\it in the opposite sense} to the
orbital spin.

\medskip
\centerline{3.1 {\sl Phenomenology of Mergers}}

In all cases we observe the appearance of a single rotating coalesced
object surrounded by a disk. The size of the disk varies greatly from
one run to another.  In some cases additional long tidal tails form as
well.  Density contour plots for runs C and D are given in Figures 1
and 2.  Logarithmic contours are plotted at intervals of 0.25 dex,
beginning at a density of 0.001 in code units (1 code unit $\equiv
2.786 \times 10^{15}$ g/cm$^3$).  The sequences shown begin just after
the onset of coalescence, times being given in code units (1 time unit
$\equiv 73 \mu$s) with $t=0$ when the two stars were 5$R_{\rm ns}$
apart.  In both cases, a single coalesced object forms on the
timescale of one orbit ({\it i.e.} $\sim$ 1ms) after the objects
collide. The system quickly becomes
axisymmetric after the two stars come into contact,
hence only a small amount of energy will be emitted
as gravitational radiation at this stage, as has been calculated in
earlier simulations by
Shibata, Nakamura \& Oohara (1992) and
by Rasio \& Shapiro (1992).
Hence our approximation of
switching off the gravitational radiation when the two stars come into
contact has been vindicated.

In run C (where the neutron stars are rapidly
rotating), far larger tidal tails develop behind the neutron stars.
These tails are clearly visible by $t=306$, and are less visible at
the same time for run D.  By a time, $t=314$, some of the material
removed from the neutron stars in the tidal tails has begun to form a
disk around the central object. As we will see shortly, this material
is shock-heated. By the end of the simulation, a large fraction of the
mass removed from the two stars remains in the disk: only some of the
material thrown off initially remains, unshocked, in the two tidal
tails. We thus observe a difference between runs C and D: in the
latter case, the lower neutron-star spin results in a delay in the
formation of tidal tails, the material has thus been ejected less far
when the two objects coalesce, and any material that has been removed
from the two neutron stars is located in the disk by the end of our
simulation.

Figure 3 illustrates the distribution of SPH particles,
in the plane of the original orbits, at the end of Run C. This plot
illustrates the three components of the final system produced in our
simulation: the central coalesced object (out to a radius $\sim 2
R_{\rm ns}$), a disk (out to a radius $\sim 6 R_{\rm ns}$), and two
elongated tidal tails, which have expanded to $\sim 20R_{\rm ns}$ by
the end of the simulation.
One has to address different questions regarding each of these
regions.
We wish to know whether the energetics of the core or the
disk are sufficient to fuel cosmological gamma-ray bursts. For this we
need to examine on one hand the thermal energies and the temperatures of the
central coalesed object and the disk as well as the mass of the disk,
and on the other hand the opacity for neutrino emission in various regions.
We are concerned with the stability and subsequent evolution of the
central object and for this we are interested in the rotational and
gravitational energies of the core.
Finally we must also estimate how
much material is ejected from the system.  This question has some
implication to the emission of gamma-ray bursts by these events and to
the possiblity that r-process nucleosynthesis is taking place in these
sites.

In Figure 4, we plot density contours of the final configuration for
run C in a plane perpendicular to that of the neutron star
trajectories. Logarithmic contours are plotted, at intervals of 0.25
dex, starting at a density of 0.0001 in code units (recalling 1 code
unit $\equiv 2.786 \times 10^{15} $ g/cm$^3$). The maximal density
decreases by about 10\%  as the two stars are tidally distorted and then
subsequently increases by about 5\% relative to the initial density
as a central coalesced object forms.  The final central density
is $\approx 5$ times the nuclear density. This is only slightly
higher than the central density of the original neutron stars.
At these
densities the polytropic index derived from the Lattimer \& Swesty
(1991) equation of state is  practically unchanged.
Figure 4 shows the
form of the central, coalesced object and the disk of material
surrounding it. It is clear from this plot that the disk is thick,
almost toroidal; the material having expanded on heating through
shocks.  This disk surrounds a central object that is somewhat
flattened due to its rapid rotation.  It is apparent from this figure
that an almost empty centrifugal funnel forms around the rotating axis
and there is almost no material above the polar caps. The funnel is
also apparent in Figure 5, where we produce a contour plot of the
integrated column density, {\it i.e.} $\int^z_\infty \rho dz$.  This
funnel might be important for the formation of gamma-ray bursts as it
provides a region in which a baryon free radiation-electron-positron
plasma could form.  Such plasma is an essential step in any
cosmological gamma-ray burst model.

\medskip
\centerline{3.2 {\sl Angular Velocity and Mass Distribution}}

For each run, we computed the ratio of system kinetic energy to the
gravitational potential energy, {\it i.e.} $T/|W|$, as a function of
time. All runs show somewhat similar behavior, with the initially more
spun-up neutron stars having slightly greater kinetic energy.  In all
runs, $T/|W|$ increased from $\sim 0.1$ to $\sim 0.17$ between $t=290$
and $t=300$. As mass was shed from the neutron stars when they produced
tails, $T/|W|$ decreased, reaching a value of $\sim 0.13$ at the end
of the simulations. For rigidly-rotating polytropes of $\gamma
\simgreat 2.2$, a secular instability occurs for $T/|W| = 0.14$.
Under such an instability, a rotationally-symmetric ellipsoid
(Maclaurin spheroid) will transform into a triaxial ellipsoid (Jacobi
ellipsoid). However, such a transformation will occur on a relatively
slow timescale (Press and Teukolsky 1973), {\it i.e.} many dynamical
timescales. Hence, even though the polytrope considered in our
simulations is sufficiently hard, and the system reaches a large
enough value of $T/|W|$, the swirling mass has resolved itself into a
central coalesced object surrounded by a thick disk {\it before} the
system has time to transform into a Jacobi ellipsoid.
A discussion concerning the subsequent stability of the central coalesced
object is given in section 4.

In Figure 6 we plot the spin angular velocities of the SPH particles
as a function of cylindrical radius. For both runs C and D, again, we
see evidence for the three regions produced in run C.  The material in
the central object ($r \simless 1.5R_{\rm ns}$) rotates rigidly; an
effect of the viscosity of our SPH fluid (which is much larger than
that of the real neutron star fluid).  In run D, the rotation curve is
somewhat broadened in the same region.  This is due to the lower spin
rates of the material within the neutron stars before they coalesced.
It is unlikely that the real neutron star fluid will be as orderly
within the coalesced object, as it is so inviscid.
For $1.5R_{\rm ns} \simless r \simless 6 R_{\rm ns}$ ({\it i.e.}\ the
disk), $\Omega_{\rm spin}$ decreases almost as $r^{-3/2}$; the slight
flattening of the power law caused by the mass contained in the disk.
For run C, the material located at greater radii is contained in the
two tails. Note that, as discussed earlier, no tails remain in run D,
and a mere few, stray particles, are found at such large distances
from the center-of-mass.  The material in the tails produced in run C
closely follows a power law, $\Omega \propto r^{-\beta}$, where $\beta
\simeq 1.8$, {\it i.e.} the rotational velocity falls off {\it faster}
than Keplerian. This may, initially, seem a rather enigmatic result,
for which we furnish the following explanation. The essential point to
realize is that the particles contained in the tails are travelling on
{\it eccentric} paths. Imagine first the simple case where all these
particles were ejected with the same angular velocities and at the
same distance from the center-of-mass, but at different times. We
would then have an ensemble of particles located at different
positions on the same ellipse -- or rather a set of identical ellipses
with different orientations of the semi-major axes in the orbital
plane.  Kepler's Second law tells us that $\Omega r^2=$ a constant for
a given ellipse, hence if our simple model were correct, we would
expect to see $\Omega \propto r^{-2}$. However, imagine now that the
particles ejected first (and now located at the largest radii) were
ejected with more angular momentum. Their ``constant'' would be
therefore {\it larger} than that for particles ejected later, and thus
now closer to the central object. This will have the effect of
softening the power law, as is observed.

In Figure 7 we plot the enclosed mass (as a fraction of total mass) as
a function of cylindrical radius for all runs.  Again, we see the
three components of the systems produced in Runs A and C: the central
object contains $\sim$ 85\% of the system mass, the disk some 13\%,
with the remaining 2\% of the material being found in the elongated
tidal tails at $r_{\rm cyl}
\simgreat 6R_{\rm ns}$.  Runs B and D show somewhat similar profiles
to those of A and C, except they leave no material in tidal tails, and
the disk is slightly less massive, containing $\sim$ 10\% of the
system mass.  For the two cases where the neutron stars were spinning
initially in the direction opposite to the orbital motion, $M_{\rm
disk} \simless 0.05 M_{\rm system}$.  The dependence of this function
on the initial neutron star spins is clear: {\it higher spin rates (in
the same sense as orbital spin) result in greater mass-loss, and more
massive disks}.

\medskip
\centerline {3.3 {\sl Temperatures and Thermal Energy}}

In Figure 8 we plot the energy per unit mass of each SPH particle
against the density as the particle site for run C (both are in code
units: units of energy/mass being $1.858 \times 10^{20}$ ergs g$^{-1}$).
We computed isotherms using Lattimer and Swesty's nuclear equation of
state, assuming nuclear matter in beta equilibrium,
and  were thus able to estimate the temperatures at the SPH-particle
locations. As stated earlier, this is an approximate method, which
is particularly sensitive for the high density regions in which the
isotherms are very near each other, that is a small change in $u$
produces a large change in $T$.  The isotherms derived from this method
are shown in Figure 9. A lot of the material around the Z-axis in the
final configuration has come from outer regions of the neutron stars,
has thus been compressed considerably, and thus heated, increasing its $T$.
Typical temperatures in the central, coalesced object are around $T
\sim 10-15$MeV. It should be noted that even if we are overestimating
the temperature of the central coalesced object by a factor of two here,
the conclusions drawn regarding the timescale for neutrino escape in
section 4.1 are not radically altered.
When material is tidally shredded from the two neutron
stars, it initially decompresses adiabatically, then subsequently
shock heats as the material from the two tidal tails come into
contact, forming a disk.
This shocked material now has densities in
the range $ 3\times 10^{11}$g cm$^{-3} \simless \rho \simless
3\times 10^{13}$g cm$^{-3}$. Again
the same polytropic index for the EOS applies as for higher densities, with
the exception of the very outermost low density layers. In the simulation,
the disk is heated by shocks to a temperature of 2-4 MeV, and contains
 $\sim 0.25 \times 10^{53}$ ergs of kinetic
energy, and  $2 \times 10^{51}$ ergs in thermal energy.

For reasonable uncertainties in the equation of state,
Lattimer and Yahil (1989) obtained a relation between gravitational
mass and binding energy of non-rotating neutron stars, which reads
$$E_{bin}=(1.5\pm 0.15) \left({{M_g} \over {M_{\odot}}}\right)^2\times
10^{53} {\rm erg}. \eqno (3)$$
If we apply it to two 1.4 M$_\odot$ neutron stars and a 2.4 M$_\odot$
central object (assuming it is stable), we find a difference of roughly
2.76 $\times 10^{53}$ erg in binding energy which would be expected to
go into internal energy (heat), as long as the system did not loose
energy by radiation yet. However, the system is not in its ground state,
but in an excited state due to its rotation. The binding energy is
decreased by the rotation energy. In the Newtonian limit we have
$E_{rot}=0.5 I \Omega^2$, with $I$ denoting the moment of inertia. In an order
of magnitude estimate we make
use of the moment of inertia for a homogeneous object
($0.4MR^2$ for a sphere and $0.5MR^2$ for a disk). The central coalesced object
is slightly elongated and we take a medium factor 0.45, a radius of
17km (cutting at $\rho=5\times 10^{13}$g cm$^{-3}$, see Figure 4), and
$\Omega=8.6\times 10^3$s$^{-1}$. This leads to $E_{rot}$=2.32$\times
10^{53}$erg or a reduced binding energy increase between two single
neutron stars and the rotating coalesced object of 4.4$\times 10^{52}$erg
or 9.2 $\times 10^{18}$ erg g$^{-1} = 9.55$ MeV per
nucleon, which should show up as additional thermal energy in the formed
compact object and is only due to stronger gravitational binding.
This agrees well with the temperature (internal energy) of the central object
of 10-15 MeV (see Figure 9).

This excercise also illustrates that the object is a relatively fast
rotator (although not at break-up and rotationally supported), which
leads to the fact that its central density is only increased by roughly
5\% over that of the initial neutron stars
(to about $1.7<10^{15}$g cm$^{-3}$).
In that regime the same polytropic index applies to the Lattimer and Swesty
(1991) EOS as at lower densities and the object is still stable.

\bigskip
\centerline{\bf 4. ADDITIONAL PHYSICAL PROCESSES}
\nobreak
\medskip

The previous sections discussed the hydrodynamical behavior during the
first few milliseconds of the merger of two neutron stars, resulting in one
central object with densities up to 10$^{15}$ g cm$^{-3}$ and a disk,
containing about 0.2-0.4M$_\odot$ of matter with densities
of 3$\times 10^{11}-1\times 10^{13}$g cm$^{-3}$.
This part of the calculation included only  (adiabatic) hydrodynamics
and a partial loss of
gravitational radiation. There are two questions which have to be asked
at this point: (i) do energy loss or energy production mechanisms exist
which work on shorter time scales, thus making the present result
invalid, and (ii) which physical processes will shape the future
behavior of that system, its stability, energy balance between nuclear energy
generation and neutrino losses, and its composition.
We address these questions below.

\medskip
\centerline{4.1 {\sl  Neutrino Escape}}

Let us first assume that all constituents are in thermal equlibrium and
that 2 MeV neutrinos populate the disk and 10 MeV neutrinos the central
object. The neutrino nucleon scattering cross section (see e.g. Tubbs
and Schramm 1975, Shapiro and Teukolsky 1983) is given by

$$\sigma_{\nu, n}={1\over 4} \sigma_0 \left({E_\nu \over m_ec^2}\right)^2
\approx \sigma_0 E_\nu^2 ({\rm MeV}) ~~~\sigma_0=1.76\times 10^{-44}{\rm cm}^2.
\eqno(4)$$

The mean free path between scattering events
$\lambda_\nu= 1/(n \sigma_{\nu,n})$ depends on the
scattering cross section $\sigma_{\nu,n}$ and
the nucleon number density $n=\rho N_A$. This can be expressed in the
following form

$$\lambda_\nu=9.9\times 10^5 \left({{10 {\rm MeV}} \over E_\nu}\right)^2
{{10^{12} {\rm g cm}^{-3}} \over \rho} {\rm \ cm}.\eqno (5)$$

Thus, for a typical density of 10$^{12}$ g cm$^{-3}$ and 2 MeV neutrinos in
the disk we find a mean free path of 246 km and for an average density of the
central object of about $8\times 10^{14}$ g cm$^{-3}$ and 10 MeV neutrinos
we find 12.4 m. In case the disk settled already to a nuclear statistical
equilibrium for that temperature and density, it will consist partially
of alpha particles (see Lamb et al. 1978), which leads however
to the same result (we have to multiply $\lambda_\nu$ by $\approx
N^2/A=4/A$, making use of the coherent scattering cross section for
nuclei and approximating the Weinberg angle by sin$^2\theta_W$=0.25).

When scattering events follow a random walk in three dimensions, the timescale
for travelling an absolute distance $d$ is
$$t(d)={3d^2 \over \lambda_\nu c}.\eqno(6)$$
It has been pointed out by Burrows and Lattimer (1986) and Burrows
(1988) that neutrinos do not follow a scattering random walk, but that their
behavior is more complex and absorptions can occur as well. Nevertheless,
equation (6) gives a good order of magnitude estimate, sufficient for our
discussion. When employing a typical height of the disk of 15 km and 10 km
for the central object, this leads to typical diffusion time scales
of $9.1\times 10^{-6}$s in the disk and $8.1\times 10^{-2}$s in the
central object.
The first number indicates that neutrino escape would already
be substantial in the disk at a time of 1ms, while this is not the case
for the central object that cools on a longer time scale. The difference
with respect to a supernova core collapse, where typical neutrino diffusion
time scales of about 1s occur, is due to the somewhat higher
temperatures obtained in the proto neutron star after stellar collapse and
bounce. With thermal energies being higher by about a factor of 2.5-3.0,
leading to a factor of 6-9 in mean free path and diffusion time scale, where
the energies enter quadratically, we have $t_{diff}=0.5-0.7$s.

\medskip
\centerline {4.2 {\sl Neutrino Emission and Cooling}}

Now we have to consider which processes produce these neutrinos and
whether their production time scales are as fast as the loss time scales.
The very high densities and high temperatures of the central object
produce neutrinos of all families essentially instantaneously in
comparison to the diffusion time scales.
Neutrinos exist in thermal equilibrium
and all neutrino families carry similar energies. The total
internal energy gain of $\approx 4.4\times 10^{52}$erg will be
released with a diffusion time scale of roughly 0.08s (as derived
above). This differs from a supernova core collapse, where the proto neutron
star releases about $2.5\times 10^{53}$erg on time scales of 1s.

Accretion onto the proto neutron star in a supernova core collapse
with a rate $\dot M$, before
the formation of the delayed shock via neutrino heating,
will increase the neutrino luminosity by (see e.g. Burrows 1988)

$$\eqalign{L_\nu  =&\dot M {GM \over R}
\left({{2} \over {1 + \sqrt{1-2GM/c^2R}}}\right)\cr
\approx & 3\times 10^{52} \left({\dot M \over 0.1M_\odot {\rm s^{-1}}}\right)
\left({M \over M_\odot}\right) \left({10{\rm km} \over R}\right)
{\rm erg~ s^{-1}}.\cr} \eqno(7)$$

A similar accretion phase, fed by matter in the disk, can occur onto the
central object in systems of merged neutron star binaries.

While in the central object energy loss due to neutrinos is limited by
the diffusion time scale controlling the possible escape, we noticed
that the neutrino diffusion timescale in the disk is very short (of the
order of $10^{-5}$s). All neutrino poduction processes occur on longer
time scales, and thus a free escape for neutrinos from these neutrino
cooling processes is guaranteed. Positron capture on neutrons and
electron capture on protons which emit anti-neutrinos and neutrinos
dominate over all other neutrinos losses (pair, photo, plasma, and
bremsstrahlung neutrinos -- Schinder et al. 1987, Itoh et al. 1989,
 1990). The rates for the disk conditions are of the order of
$10^{20}-10^{22}$erg g$^{-1}$s$^{-1}$ (Fuller et al 1980,1982).
The next largest energy loss rate due to pair and plasma neutrinos amounts
only to about $10^{15}-10^{18}$erg g$^{-1}$s$^{-1}$ (Itoh et al. 1989, 1990).
The time scale for electron and positron captures (and accompanying neutrino
production) is of the order $10^{-3}-10^{-4}$s, which underlines our previous
conclusion of free escape (with diffusion time scales of the order of
$10^{-5}$s), i.e. the full energy loss is encountered and no trapping occurs.

\medskip
\centerline {4.3 {\sl Thermonuclear Energy Generation in the Disk}}

In addition to the cooling processes, we have to consider nuclear
energy generation in the disk. Simple nuclear network calculations
show that any initial protons combine with neutrons to form alpha
particles on a time scale of less than $10^{-17}$s for the density and
temperature conditions in the disk. This releases an energy of 7.07
MeV per nucleon or $6.8\times 10^{18}$erg g$^{-1}$ or $1.36\times
10^{51}$ erg per 0.1M$_\odot$ of disk matter burned to helium.  Disk
material in beta equilibrium would consist of less than 1\% protons.
Thus, a disk in weak equilibrium would burn 2\% of its mass (equal
amounts of neutrons and protons) instantaneously, releasing about
$10^{50}$erg (on a $10^{-17}$s time scale). This would heat and
somewhat expand the disk because of a higher energy generation than
neutrino cooling rate.

The subsequent evolution is complex and depends on many details, i.e.
the resulting density and temperature of the disk and the free proton
fraction $Y_p$. High densities (large electron Fermi energies) and/or
temperatures favor electron captures on protons. For the disk
densities  electron capture time scales of
about $10^{-1}-10^{-4}$s with a neutrino cooling rate close to
$10^{20}-10^{24}$erg g$^{-1}$s$^{-1}$ are obtained
in the temperature range of 1-10MeV. The positron capture rates on neutrons
are generally lower, because positrons have a negative Fermi energy.
Similar values as for electron captures on protons can only be attained
at high temperatures and lower densities.
These rates would usually balance in order to
keep beta equilibrium. But each proton produced will burn to helium on
much shorter time scales, thus dominating over electron captures.
The depletion in the free proton abundance $Y_p$ is replenished
by further positron captures on neutrons, and gradually the free
neutrons would burn to He via conversion to protons.

A total energy gain would occur if the density and proton abundance
(determining the electron Fermi energy and electron capture rate on
protons) become small enough that neutrino losses from electron
captures are negligible, and the temperatures are low enough that the
neutrino losses in each positron capture reaction (producing one
proton) become smaller than 14MeV, which is 1/2 of the energy released when
two resulting protons burn with two available neutrons to He. This
is the case for temperatures of $\le$1MeV, $\rho Y_e\le 10^9$g
cm$^{-3}$, and $Y_p\le 10^{-3}$ (tables 2 and 3 in Fuller et al. 1982
and private communication). The low temperatures are required as the
weak capture cross sections scale with $E^2$ and thus the average
energy of the positron and the escaping neutrino are higher than the
thermal value. For the conditions discussed above the time scale of
burning neutrons via protons to He is of the order 100s.

On the other hand the low densities ensure electron Fermi energies of
less than 5MeV. Thus, the formation of heavier elements via neutron
captures and beta decays with typical Q-values of 13-15MeV will not be
inhibited by electron captures on neutron-rich nuclei. An r-process
similar to the r-process suggested for inhomogeneous big bang
nucleosynthesis (Applegate, Hogan, Scherrer 1988; Thielemann et al.
1991) with light nuclei embedded in a neutron bath will occur. Then
the long time scales required to convert neutrons via protons to
helium can be overcome by this feature of the burning process.  Once
the first heavy seeds are available, neutron captures and beta decays
(not temperature dependent) will drive the burning on beta decay time
scales of 0.1 to 0.01s and cause a net energy deposition in the low
density and temperature parts of the disk. The maximum energy of
photons emitted in the nuclear reactions would result from beta decays
far from stability with decay energies up to 15-20 MeV.  The scenario
is very similar to that of ``cold decompression'' (Lattimer et al.
1977, Meyer 1989), with the exception that no energy loss terms and an
adiabatic expansion fueled by the energy generation was considered in
the latter.  The total expected energy release is of the order
$f\times (5-6)\times 10^{51}$ erg, dependent on the distribution of
nuclei produced and the fraction $f$ of the disk involved.

In the interior of the colliding neutron stars one finds 5-30\%
protons, depending on the equation of state (Weber and Weigel 1989).
This increase is due to the fact that neutrons and protons are also
degenerate at such high densities and the neutron and proton Fermi
energies are of importance as well. If such matter would be ejected
during the collision on shorter than weak equilibrium time scales, up
to 60\% of the disk matter could burn instantanesously, releasing
about $3\times 10^{51}$erg in the disk. Our calulations show that the
ejection of matter into the disk occurs on time scales comparable to
weak equilibrium time scales and a behavior closer to the earlier
discussion is expected.

In none of the cases would the energy release be sufficient to unbind
and ``evaporate'' the whole disk, which has a total binding energy of
$\sim 4 \times 10^{52}$erg but possibly the outer lower density parts
can escape. Further calculations which treat energy loss rates and
thermonuclear reactions consistently as part of the hydro code will
hopefully be able to answer the open ``details''.

\medskip
\centerline {4.4 {\sl Stability of the Central Coalesced Object}}

Our hydro calculations produced a central object which was stable on
the time scales considered in this calculation. We have to consider
several points to understand that outcome and its generality.
The main question concerns the limiting neutron star mass for a given
equation of state. The object is rotating slower than, but close to the,
Kepler frequency of typically $\Omega_K=10^4$s$^{-1}$.
This allows an increase beyond the limiting mass of
non-rotating neutron stars of 14-17\% (Friedman, Ipser, Parker 1986;
Friedman and Ipser 1987; Weber and Glendenning 1992), the largest
values are related to the softest equations of state.

The core is rapidly rotating but $Jc/GM^2 \approx .6 < 1$. (similar
results were obtained by Shibata Nakamura \& Oohara (1992) and by Rasio
\& Shapiro (1992)).  Thus the core is not supported completely by rotation.
This can be understood intuitively from the
following argument. The angular momentum just before the stars are
in contact is
$$
J \approx 2 M \Omega R^2 = 2 M (G M /4 R^3)^{1/2} R^2 = (G M^3 R )^{1/2}.
\eqno(8)
$$
The ratio $J/ (4 G M^2 /c)$ determines whether the object is
rotationally supported. We find
$$
{J c \over 4 G M^2 } = { 1 \over 4 } \left({R \over r_g(M)}\right)^{1/2},
\eqno(9)
$$
where $r_g(M)$ is the gravitational radius of one of the stars,
i.e. $r_g \approx 2 (M/1.4 M_\odot)$km. With $R \approx 10$km
we have that $R/r_g \approx 5$ and hence $Jc/GM^2 \approx .5$.
The ratio is slightly higher in our case because of the spin of
the neutrons stars. However, it is quite impossible to reach
$Jc/GM^2 > 1$. The intuitive explanation for the phenomenon is that
while orbiting each star is attracted only by its companion and the
centrifugal force has to balance this gravitational attraction.
However, once the two objects coalesce the gravitational
force towards the center results from a total mass of $2 M$ which
is too large to be stopped purely by the rotation.

The behavior of the equation of state at high densities is dominated by
the compressibility $K$ and the possible appearance of new degrees of
freedom, i.e. existence of excited baryons (hyperons) in addition to
neutrons and protons, kaon condensates, etc. If it is permitted to
populate these states rather than neutrons and protons with their huge
Fermi energies, a smaller pressure and softer EOS results, which causes a
smaller limiting neutron-star mass. Our polytropic index was adjusted to
the Lattimer and Swesty EOS with a compressibility $K=180$ MeV.
Only neutrons, protons, and leptons are considered at high densities. This
corresponds to a large limiting neutron star mass of 2.2M$_\odot$.

Most recent relativistic equations of state result in limiting masses
for non-rotating neutron stars below 2.2M$_\odot$, down to
1.4-1.6M$_\odot$ (Glendenning et al. 1992).
With a 14-17\% increase for objects close to their
Kepler frequencies, we would start out with stable central objects for
the upper end of this mass range. We would expect neutrino release
from the hot object and mass accretion until
accretion, cooling or loss of angular momentum lead to surpassing
the limiting mass and cause the collapse to a black hole.
We expect that this will take place on a time scale of seconds.
A limiting non-rotating neutron star mass smaller than 2M$_\odot$ would
lead to an unstable coalesced object from the time of its formation.
Thus, a black hole would form after about 1ms and a neutrino
burst of only a few times $10^{50}$erg would result during the collapse
(Gourgoulhon and Haensel 1993). This will be
enhanced by continuing accretion from the disk.

This clearly shows that the outcome of a neutron star merger, its
neutrino signal, electromagnetic radiation, and probably also the disk behavior
depend decisively on the limiting neutron star mass. If identified with
observed events (i.e. a gamma ray burst), we could finally expect
observational contraints for this theoretically quite uncertain
quantity.

\bigskip
\centerline {5. DISCUSSION}
\nobreak
\medskip

For all the neutron-star spins considered here, the merger results in
a rapidly rotating, though not completely rotationally supported,
central coalesced core surrounded by a disk of ejected material.
A few percent of the total mass is ejected
into long extended tails in some of the cases.  During the merger
event material from the two neutron stars is compressed adiabatically,
increasing the temperature of the gas to $\approx$10 MeV and it
becomes a strong source of neutrinos.  The neutrino cooling time is
around $10-100$ms. The hot rapidly rotating core is on the edge of
stability and its subsequent evolution depends on the true equation of
state.  The disk material gets heated via shocks and its temperature
is $\approx 5$MeV. Since its density is much lower its neutrino cooling
time is significantly shorter. Thermonuclear reactions can, however,
heat the disk and its thermal evolution depends on a delicate balance
between heating and cooling.

Our calculations included several simplifying approximations and it is
worth while to examine their validity before
turning to the interpretation of the results: (i) We employed a
polytropic equation of state instead of a realistic one.  However we
found that by the end of the calculations the central density
increased only by $5\%$ and the polytropic approximation
to the Lattimer \& Swesty
(1991) equation of state is valid at this density range. (ii) We
did not include neutrino transport and cooling. The neutrino cooling
time of the core ($10-100$ms) is sufficiently long that
there are no significant neutrino losses
up to the moment that we stop the calculations.
The time scale for neutrino cooling of the disk is,
however, much shorter and our approximation breaks down there.  (iii)
We switch off the gravitational radiation backreaction force when
the two neutron stars touch each other.  The core
becomes axisymmetric within a fraction of a millisecond after the two
stars have come in contact, and only a small amount of energy is
released via gravitational radiation during this phase (Shibata
\etal, 1993, Rasio \& Shapiro 1993).
(iv) Our worst approximation is that apart from
gravitational radiation backreaction we did not include
other general relativistic effects,
There is no physical justification for
that, only a technical one, the required
three dimensional hydrodynamics calculations
are beyond the current scope of numerical relativity.
General relativistic effects are the major uncertainty in the
calculations that we present here. We have no way of estimating them
quantitatively. Qualitatively we expect that they might enhance the
dynamical instability of the central rotating core.

The mass of the central object ranges between $2.4 - 2.6 M_\odot$.
This is above most upper limits for cold neutron star masses even with
the stiffest equations of state, (e.g.
the limiting mass is 2.2M$_\odot$ for non-rotating neutron star with
the Lattimer-Swesty equation of state that we follow).
It is clear that if the real equation of state is
soft, the central object must be unstable and it will collapse quickly
to a black hole.
The situation is less clear with a hard equation of
state.
The maximum mass allowed increases by up to 17\% when the neutron star
is rotating rapidly.
Furthermore, our hydrodynamics
calculations show that the Newtonian system with a hard equation of
state, seems stable and does not show any tendency to collapse.
At the end of the calculation
the central polytropic index is large which
indicates that the system is far from the instability that sets in as
the polytropic index decreases to 4/3.
Here it is possible that the combination of rotation and
thermal pressure may prevent the direct collapse.
The collapse might be induced
later when additional mass from the disk accretes onto the central
object or when it cools down.  General relativistic effects, which we
ignore here, might have a critical effect and cause the central merged
object to collapse sooner. As the central core is on the edge of
stability it might be the case that in some cases (depending e.g.  on the
spin angular momentum) the core remains extended initially while in
others it collapses directly to a black hole.  In either case the
ultimate final configuration is a black hole.

The total neutrino emission depends on how fast the core collapses to
a black hole. One can expect appreciable neutrino emission if the
rotating configuration is stable, or if the combination of rotational
and thermal energy make the system stable, halts the collapse until
the core cools or accretes additional matter.  If general-relativistic
effects or a softer equation of state will make the system unstable a
black hole will form after a few msec without much neutrino emission
form the core.

The disk extends up to 60km or so.  Its total mass is $\approx
0.4M_\odot$, and its total kinetic energy $\approx 2.5 \times 10^{53}$
erg.  From our SPH runs, we see that the disk contains $\sim 10^{51}$
erg of thermal energy, with some of the material being shock-heated up
to temperatures of $\sim$ 2-4 MeV.  The disk is optically thin to
neutrinos, and neutrino cooling exceeds heating for temperatures
$T>$1MeV, densities $\rho Y_e>10^9$g cm$^{-3}$ and free proton
fraction $Y_p>10^{-3}$.  Hence we expect neutrino cooling to lead to a
flattened disk in the inner high density part.

There occurs a thermonuclear flash of burning of the initial protons (with
neutrons) in the disk to He. At locations in the disk where after that
initial flash $T$, $\rho Y_e$ ,and $Y_p$ are equal or smaller than the
values listed in the previous paragraph net energy deposition rather
than neutrino losses will dominate and a further expansion is expected.
The low densities ensure electron Fermi energies less than 5MeV. Thus,
the build-up of heavy elements from the seed nuclei, produced in the
early thermonuclear flash, via neutron captures and beta decays (with
Q-values of typically 13-15MeV) is not inhibited by reverse electron
captures on these neutron-rich nuclei.
If a fraction $f$ of the whole disk is burned to heavy elements,
this would result in a total release of $f\times (5-6)\times 10^{51}$ erg.
We expect some of the material of the outer disk and extended arms to
become unbound from the system due to this nuclear energy release
and to contain r-process matter produced during the decompression phase.

Ratnatunga \& van den Bergh (1989) find core collapse supernova rate
in our galaxy of $2.2\pm1.1$ per century,
10$^{-6}$ M$_{\odot}$ to 10$^{-4}$ M$_{\odot}$ of r-process
material must be ejected per supernova event (Cowan, Thielemann
\& Truran 1991) to produce the observed abundance of r-process material.
Eichler \etal (1989) suggested that, alternatively, the  r-process takes place
in neutron star mergers. They estimated the production rate using
the merger rate estimated by
Clark  van den Heuvel \& Sutantyo. Recently,
Narayan \etal (1991) and Phinney (1991)
estimate that there are $10^{-5.5 \pm .5}$ neutron star mergers per
year per galaxy. Using the updated rate we
find that  to explain r-process nucleosynthesis as
a by product of neutron star merger one requires the ejection of
$10^{-2} M_\odot$ to $1M_\odot$ per event. The upper limit is clearly
impossible, however, the lower range is of the same order of magnitude
of what we observed in the long tails in some of the mergers.
Some of the r-process material is observed in very low metalicity stars
and galactic evolution models would require that it is produced as early as
${\rm a\ few} \times
10^7-10^8$years after the formation of the galaxy. This is comparable
with the life time of the famous binary pulsar PSR-1913+16 and one might
expect that there is no problem caused by this time scale.
This suggests that some r-process material could be produced via this
mechanism even on such a short time scale.

We turn now to the implications to $\gamma$-ray bursts models.
Cosmological $\gamma$-ray bursts require $\approx 10^{51}$ ergs with a
rise time as short as tens of msec (for some of the short bursts).
The hot coalesced core is one possible source of this energy.  If the
core does not collapse directly to a black hole it will
emit its thermal energy as neutrinos. The neutrino flux is sufficiently
large that $\approx 10^{-2}$ to $10^{-3}$ of it could be converted to
electron-positron  pairs via $\nu \bar \nu \rightarrow e^+ e^-$ (Goodman \etal,
1987) and those could produce a $\gamma$-ray burst (Eichler \etal, 1989).
The time scale for the neutrino burst is short enough to accommodate even
the shortest rise times observed. Additional neutrino luminosity
could arise on a longer time scale
from accretion of the disk material on the central object.

An additional energy source that could power a $\gamma$-ray burst from
a neutron star merger is the disk surrounding the central object. This
energy source can operate regardless of the question of whether the
central object collapse directly to a black hole or not. In this case
the neutrino emission is not sufficient to power a burst via
neutrino annihilation, however, Narayan \etal (1992) have suggested
that the magnetic energy density in the disk can reach
equipartition with the kinetic energy (corresponding to $10^{16}$
gauss) and recombination of such a magnetic field could produce an
electromagnetic burst. It is not clear what will be the time scale for
the emission from this disk. It could take several seconds or even
longer, depending on the viscosity in the disk, which is unknown at
present. As was stressed by Narayan \etal (1992) the large variability
in observed properties of $\gamma$-ray bursts could be explained by
those two alternative energy sources. An additional source of diversity,
which we discover here, is the distinction between systems that collapse
directly to a black hole and those that undergo a longer rotating core phase.

The initial $\gamma$ rays that are produced are not observed directly.
The huge optical depth of the resulting electron-positron plasma would
result in an optically thick fireball (Goodman, 1986; Shemi \& Piran,
1990; Paczynski, 1990; Piran \& Shemi, 1993).  The observed
$\gamma$-rays emerge much later when the fireball becomes optically
thin (due to its rapid expansion) or when it interacts with
interstellar material (Meszaros \& Rees, 1993). This late stage
determines both the duration and the spectrum of the bursts.
Shemi \& Piran (1990) have shown that
to produce a $\gamma$-ray burst the fireball must be practically free
from baryonic load. If more than $10^{-5}M_\odot$ baryons are injected
into the fireball they will prevent the formation of a $\gamma$-ray
burst.  Larger amounts of mass are ejected
in a neutron star merger.  However  all the mass is ejected into the
equatorial plane.  In all mergers simulated, we note the the regions
above the poles of the coalesced object are relatively free of
material.  The accuracy of our simulation is of course not enough to
estimate the amount of matter in those funnels, but it seems that
those funnels may provide an avenue for neutrino escape and the
production of a clean radiation fireball. This suggests that the
$\gamma$-rays are beamed and appear only in certain directions.

Acknowledgements

We would like to thank Lars Bildsten, Adam Burrows, Joan Centrella,
Curt Cutler, Dalia Goldwirth, Chris Kochaneck, Jim Lattimer,
Ramesh Narayan, Doug Swesty and Fridolin Weber for helpful discussions.
We are particularly grateful to D. Swesty and J. Lattimer for making available
their nuclear equation equation of state code.
This research was
supported in part by NASA grant NAG 5-1904, NSF grants AST 89-13799
and AST-9206378, and a
BSF grant to the Hebrew University. M.B.D. gratefully acknowledges
the support of an R.\ C. Tolman Fellowship at Caltech.

\vfill\eject

\def\etal{{\it et. al.}}
\centerline{REFERENCES}
\medskip

\def\ref{\par \smallskip \noindent \hangindent .5in \hangafter 1}
\ref
Abramovici, W.E. \etal$\,$  1992, {\rm Science}, {\rm 256}, 325.
\ref
Applegate, J.H., Hogan, C.J., Scherrer, R.J. 1988,
{\rm ApJ}, {\rm 329}, 572
\ref
Blanchet, L. Damour, T. \& Schafer, G. 1990, MNRAS, {\rm 242}, 289
\ref
Bildsten, L., and Cutler, C. 1992, ApJ, 400, 175
\ref
Burrows, A. 1988, {\rm ApJ}, {\rm 334}, 891
\ref
Burrows, A., Lattimer, J.M. 1986, {\rm ApJ}, {\rm 307}, 178
\ref
Clarke, J. P. A, van den Heuvel, E. P. J., and
Sutantyo, W. 1979, {A \& A}, {\rm 72}, 120.
\ref
Cowan, J., J.,  Thielemann, F., K., \&  Truran J. W.,
 1991 {\rm Phys. Rep.}, {\rm 208}, 267
\ref
Cutler, C., \etal$\,$ 1993, Phys. Rev. Lett., {\rm 70}, 2984.
\ref
Davies, M.\ B., Benz, W., and Hills, J.\ G., 1992, ApJ, 401, 246
\ref
Eichler, D., Livio, M., Piran, T., and Schramm, D. N. 1989,
Nature, {\rm 340}, 126
\ref
Friedman, J.L., Ipser, J.R., Parker, L. 1986, {\rm ApJ}, {\rm 304}, 115
\ref
Friedman, J.L., Ipser, J.R. 1987, {\rm ApJ}, {\rm 314}, 594
\ref
Fuller, G.M., Fowler, W.A., Newman, M. 1980, {\rm ApJS}, {\rm 42},
447
\ref
Fuller, G.M., Fowler, W.A., Newman, M. 1982, {\rm ApJS}, {\rm 48},
279
\ref
Glendenning, N.K., Weber, F., Moszkowski, S.A. 1992,
{\rm Phys. Rev.} {\rm C45}, 844
\ref
Goodman, J., 1986, ApJ, {\rm 308}, L47
\ref
Goodman, J., Dar, A. and Nussinov, S. 1987, ApJ,
{\rm 314}, L7
\ref
Gourgoulhon, E., Haensel, P. 1993, {\rm A \& A} {\rm 271}, 187
\ref
Itoh, N., Adachi, T., Nakagawa, M., Kohyama, Y., Munakata, H. 1989,
{\rm ApJ}, {\rm 339}, 354
\ref
Itoh, N., Adachi, T., Nakagawa, M., Kohyama, Y., Munakata, H.
1990,{\rm ApJ}, {\rm 360}, 741
\ref
Kidder, L. E., Will, C. M., and Wiseman, A. G. 1993,
Phys. Rev. D, {\rm 47}, 3281
\ref
Kochanek, C., 1992, ApJ, 398, 234
\ref
Lai, D., Rasio, F. A., and Shapiro, S. L. 1993, ApJ, {\rm 406}, L63
\ref
Lamb, D.Q., Lattimer, J.M., Pethick, C.J., Ravenhall, D.G. 1978,
{\rm Phys. Rev. Lett.} {\rm 41}, 1623
\ref
Lattimer, J.M., Mackie, F., Ravenhall, D.G., Schramm, D.N. 1977,
{\rm ApJ} {\rm 213}, 225
\ref
Lattimer, J.M., Swesty, F.D. 1991, {\rm Nucl. Phys.} {\rm A}, 535, 331
\ref
Lattimer, J.M., Yahil, A. 1989, {\rm ApJ} {\rm 340}, 426
\ref
Lincoln, C. W. \& Will, C. M. 1990, Phys. Rev. {\rm 42}, 1123
\ref
Meszaros P., \& Rees, M. J., 1993, ApJ, 405, 278
\ref
Meyer, B.S., 1989, {\rm ApJ} {\rm 343}, 254
\ref
Nakamura, T. \& Oohara, K. 1989, Prog. Theo. Phys. {\rm 82}, 1066
\ref
Nakamura, T. \& Oohara, K. 1991, Prog. Theo. Phys. {\rm 86}, 73
\ref
Narayan, R., Paczy\'nski, B., and Piran, T. 1992, ApJ,
{\rm 395}, L83
\ref
Narayan, R., Piran, T. and Shemi, A., 1991, ApJ, {\rm 379},
L17
\ref
Oohara, K. \& Nakamura, T , 1989, Prog. Theo. Phys. {\rm 82}, 535
\ref
Oohara, K. \& Nakamura, T , 1990, Prog. Theo. Phys. {\rm 83}, 906
\ref
Paczy\'nski, B., 1990, ApJ, {\rm 363}, 218
\ref
Paczy\'nski, B., 1991, Acta Astronomica, {\rm 41}, 257
\ref
Phinney, E. S., 1991, ApJ, {\rm 380}, L17
\ref
Piran, T., 1990, in Wheeler, J. C., Piran, T. and Weinberg, S.
{\rm  Supernovae} World Scientific Publications
\ref
Piran, T., Narayan, R. and Shemi, A., 1992, in
{\rm  Gamma-Ray Burst, Huntsville, 1991}
Paciesas W. S. and  Fishman, G. J., eds, AIP press
\ref
Piran, T. \& Shemi, A., 1993, ApJ, {\rm 403}, L67
\ref
Rasio, F. A. \& Shapiro, S., L., 1992, ApJ, {\rm 401}, 226
\ref
Ratnatunga K.,  \& van den Bergh, S., 1989, ApJ, {\rm 343}, 713
\ref
Shemi, A. and Piran, T., 1990, ApJ, {\rm 65}, L55
\ref
Schinder, P.J., Schramm, D.N., Wiita, P.J., Margolis, S.H., Tubbs,
D.L. 1987 , {\rm ApJ}, {\rm 313}, 531
\ref
Schutz, B.F., 1986, {\rm Nature}, {\rm 323}, 310
\ref
Shapiro, S.L., Teukolsky, S.A. 1983,
{\rm Black Holes, White Dwarfs, and Neutron Stars}, (Wiley, New York)
\ref
Shibata, M., Nakamura, T., \& Oohara, K.,
1992, Prog. Theo. Phys. {\rm 88}, 1079
\ref
Thielemann, F.-K., Applegate, J., Cowan, J.J., Wiescher, M. 1991,
in {\rm Nuclei in the Cosmos}, ed. H. Oberhummer
(Springer-Verlag), p.147
\ref
Thorne, K.S., 1987, {\rm 300 Years of Gravitation}, S.W. Hawking \&
  W. Isreal, eds., (Cambridge Univ. Press: Cambridge) 378
\ref
Tubbs, D.L., Schramm, D.N. 1975, {\rm ApJ}, {\rm 201}, 467
\ref
Weber, F., Weigel, M.K. 1989, {\rm Nucl. Phys.}, {\rm A505}, 779
\ref
Weber, F., Glendenning, N.K., Weigel, M.K. 1991, {\rm ApJ}, {\rm 373}, 579

\vfill\eject

\midinsert
  \line \bgroup \hss
  $$\vbox{
\halign{ # \hfill \quad & # \hfill \quad & \hfill # \quad & # \hfill
 \quad & # \hfill \cr
  \multispan5 \hfill {Table I} \hfill \cr
  \omit & \omit & \omit \cr
  \multispan5 \hfill {Characteristics of Runs} \hfill \cr
\noalign{\vskip 5pt}
\noalign{\hrule}
\noalign{\vskip 3pt}
\noalign{\hrule}
\noalign{\vskip 10pt}
Run & $\Omega_{\rm contact}$ & Npart & Reallign &
$M_{\rm disk}/M_{\rm system}$ \cr
\noalign{\vskip 5pt}
\noalign{\hrule}
\noalign{\vskip 10pt}
A & 0.2329 & 8542 & no & 0.13 \cr
B & 0.0478 & 8542 & no & 0.10 \cr
C & 0.2143 & 8542 & yes & 0.13 \cr
D & 0.0109 & 8542 & yes & 0.10 \cr
\noalign{\vskip 10pt}
E & -0.0439 & 2512 & yes & 0.05 \cr
F & -0.1257 & 2512 & yes & 0.05 \cr
\noalign{\vskip 5pt}
\noalign{\hrule}
\noalign{\vskip 10pt} }}$$
  \hss \egroup
\endinsert
\vskip .5in

\vfill\eject

%\magnification 1200
%\hoffset 0.5truein
%\hsize 6truein
%\nopagenumbers
\centerline{\bf FIGURE CAPTIONS}
\bigskip
\noindent {\bf Figure 1:}
Density contour plots for run C, in the plane
of the neutron star trajectories. Time and distances are given in code units
(1 time unit $\equiv 73 \mu$s, 1 distance unit $\equiv R_{\rm ns} = $10km).
Logarithmic contours are plotted, at intervals of 0.25 dex, beginning
at a density of 0.001 in code units (1 code unit $\equiv 2.786
 \times 10^{15} $g/cm$^3$).

\noindent {\bf Figure 2:}
Density contour plots for run D. Units as given in Figure 1.

\noindent {\bf Figure 3:}
The distribution of SPH particles, in the plane of the original orbits,
 at the end of run C. Distance units as given in Figure 1.

\noindent {\bf Figure 4:}
Density contour plot of the final configuration in run C in a plane
perpendicular to that of the neutron star trajectories.  Logarithmic contours
are plotted, at intervals of 0.25 dex, starting at a density of
0.0001 in code units (1 code unit $\equiv 2.786 \times 10^{15}$ g/cm$^3$).

\noindent {\bf Figure 5:}
Contour plot of column densities, {\it i.e.} at a given location ($r, z$),
column density, $\rho_{\rm c} = \int^z_\infty \rho dz$.

\noindent {\bf Figure 6:}
The spin angular velocities of the SPH particles as a function of
cylindrical radius. For run C (figure a) and run D (figure b). Both
angular velocity and distance are in code units
(1 distance unit $\equiv R_{\rm ns} = $10km, $\Omega =1$ is equivalent
to a spin period of 0.461ms).

\noindent {\bf Figure 7:}
The enclosed mass (as a fraction of total mass) as a function of
cylindrical radius (in units of the neutron-star radius).

\noindent {\bf Figure 8:}
The energy per unit mass of each SPH particle  against the density
as the particle site for the final configuration in run C.
Both are given in code units, energy/mass
 units  being $1.858 \times 10^{20}$ ergs/gm.

\noindent {\bf Figure 9:}
Isotherms produced using Lattimer and Swesty's nuclear equation of
state, assuming beta equilibrium. Isotherms are draw for
$T=$1.75, 2, 3, 5, 7, 10, 15, and 25MeV.

\bye